\documentclass[12pt,preprint]{article}
\usepackage{amsmath}
\usepackage{graphics}
\usepackage{epsfig}
\renewcommand{\vec}[1]{{\bf #1}}
\usepackage{bbm}
\newcommand{\UM}{\mathbbm 1}
\newcommand{\eqb}{\begin{equation}}
\newcommand{\eqe}{\end{equation}}
\newcommand{\dmb}{\begin{displaymath}}
\newcommand{\dme}{\end{displaymath}}
\newcommand{\pd}{\partial}

\newcommand{\eab}{\begin{eqnarray}}
\newcommand{\eae}{\end{eqnarray}}

\newcommand{\be}{\begin{equation}}
\newcommand{\ee}{\end{equation}}

\begin{document}
\begin{titlepage}
\begin{flushright} 
\end{flushright}
\vspace{0.6cm}

\begin{center}
\Large{Loop expansion in Yang-Mills thermodynamics}

\vspace{1.5cm}

\large{Ralf Hofmann}

\end{center}
\vspace{1.5cm} 

\begin{center}
Institut f\"ur Theoretische Physik\\ 
Universit\"at Heidelberg\\ 
Philosophenweg 16\\ 
69120 Heidelberg, Germany
\end{center}
\vspace{1.5cm}
\begin{abstract}

We argue that a selfconsistent spatial coarse-graining, 
which involves interacting (anti)calorons of unit 
topological charge modulus, implies that real-time loop expansions of thermodynamical 
quantities in the deconfining phase of SU(2) and SU(3) 
Yang-Mills thermodynamics are, modulo 1PI resummations, determined 
by a finite number of connected bubble diagrams.
     . 
 
\end{abstract} 

\end{titlepage}

\section{Introduction and mini-review}

A reliable approximation of the high-temperature thermodynamics related to four-dimensional 
pure Yang-Mills theories in terms of a small-coupling expansion is impossible\footnote{We restrict our 
discussion to the gauge group SU(2). 
Larger groups can, in principle, be investigated 
by the identification of possible SU(2) embeddings. In fact, our results of Secs.\,2 and 3 are also valid for 
SU(3), compare with \cite{Hofmann2005}. However, for 
SU(N), $\mbox{N}\ge 4$, the phase diagram is not unique and hence the concept of just one 
deconfining phase is false\cite{Hofmann2005}.}\cite{Linde1980}. 
The nonconvergence of the small-coupling expansion is tied to the fact that 
a too naive a priori estimate  -- an empty (trivial) ground state -- is invoked to 
construct an approximating series for the full partition function. Recall that fluctuations of 
nontrivial topology, having a profound impact on the ground-state estimate, 
are completely ignored in small-coupling expansions because their weight 
possesses an essential zero at vanishing coupling. As a consequence, 
the strongly correlating effects of these extended field 
configurations \cite{Polyakov1975} are completely ignored: A fact 
which is expressed by the tree-level masslessness and only 
weak radiative screenings of all gauge bosons leading to the 
nonconvergence of the expansion. Loosely speaking, the physical expansion parameter, which 
is not the small coupling constant but the ratio of the typical action of a 
quantum fluctuation to $\hbar$, is not guaranteed to be small due to the 
unconstrained dynamics of massless gauge bosons.  

The purpose of the present work is to provide 
arguments on why an a priori estimate for the ground state of an SU(2) 
Yang-Mills theory at high temperatures (deconfining phase\footnote{We refrain here from discussing in detail the other 
two phases, preconfining and deconfining, 
see \cite{Hofmann2005}. While thermodynamical quantities 
are one-loop exact in the preconfining phase there exist 
asymptotic expansions in the confining phase, see for example \cite{GiacosaHofmannSchwarz2006}.}), 
which is obtained by a selfconsistent and 
sufficiently local spatial coarse graining\footnote{This refers to the fact that 
only configurations with topological charge modulus 
$|Q|=0,1$ need to be taken into account, see below.} over interacting and 
stable BPS saturated topological field configurations, 
leads to a rapidly converging loop expansion. This claim rests on the 
selfconsistent emergence of a temperature-dependent scale of maximal resolution which also 
generates a mass gap on tree-level (adjoint Higgs mechanism) for the two off-Cartan modes 
in unitary gauge (ultraviolet and infrared cutoff). Both effects imply 
that the typical action of residual quantum fluctuations in the effective theory 
is small.   

Before we start the present discussion we consider it helpful to remind the reader 
of the dynamical situation leading to the emergence of a 
highly nonperturbative ground state even at 
large temperatures, $T\gg\Lambda$. Here $\Lambda$ denotes the 
Yang-Mills scale \cite{Hofmann2005,HerbstHofmann2004}. 
Stable, that is, BPS saturated topological defects of 
a trivial-holonomy (or Harrington-Shepard (HS) (anti)calorons with $|Q|=1$) \cite{HS1977}) 
interact by large-scale (compared to the scale parameter $\rho$) gluon exchanges. This generates dynamical magnetic 
substructure in the (anti)calorons whose motion \cite{Forkel2005} is determined by 
short-scale gluon fluctuations\footnote{Nontrivial holonomy is associated with 
a mass scale $\propto T$ expressing itself 
in singular gauge by the Polyakov loop at spatial infinity 
not coinciding with a member of the SU(2) center $-\UM,\UM$. 
The $A_4$-component of the (anti)caloron configuration then effectively 
serves as an adjoint Higgs field. As a consequence, a 
BPS monopole and its antimonopole emerge which are spatially separated if 
$\rho$ does not vanish \cite{Nahm1984,LeeLu1998,KraanVanBaalNPB1998}. By computing 
the one-loop quantum weight for a nontrivial-holonomy SU(2) caloron it 
was shown in \cite{Diakonov2004} that 
a small (anti)caloron holonomy induces attraction between the monopole and 
its antimonopole while a large holonomy yields repulsion. While the likelihood of the former 
situation is determined by the quantum weight 
of a HS caloron and thus, depending on $\rho$ and $T$, can be 
of order unity \cite{GrossPisarskiYaffe1981} the probability for 
repulsion is estimated by a Boltzmann factor 
$\sim e^{-40}$ \cite{Hofmann2005}: The by-far dominating situation is monopole-antimonopole attraction which drives the (anti)caloron back 
to trivial holonomy (monopole-antimonopole 
annihilation). The rare process of (anti)caloron 
dissociation by monopole-antimonopole repulsion generates isolated 
(anti)monopoles whose magnetic charge is screened by 
intermediate small-holonomy (anti)calorons. Screened (anti)monopoles 
do not contribute to thermodynamical quantities like the 
pressure, the energy density, and 
the entropy density: They are too massive and too rare. Despite the fact 
that there is an extremely small ratio of the number of stable 
and and the number of to be annihilated monopoles at any 
instant of time and 
at any temperature ($T\gg \Lambda$) the absolute 
density of stable and screened monopoles 
increases with temperature \cite{Hofmann2005}. This leads 
to the lattice-observed phenomenon of a 
spatial string tension rising 
as $T^2$ \cite{GiovannangeliKorthals Altes2001}, see \cite{Polyakov} for the 
deep theoretical reasons. An according microscopic model was constructed in 
\cite{MullerPreussker2006}. This model builds on the existence of a typical 
spatial volume inhabited by a dissociated, large-holonomy (anti)caloron 
with $|Q|=1$ \cite{Hofmann2005}.}. At the same time the presence of (anti)calorons 
induces consecutive scatterings of 
all those gluons that are sufficiently close to 
their mass shell to propagate over large distances. Upon a selfconsistent spatial coarse-graining 
the former situation identifies ground-state pressure and  ground-state energy density with that of 
a linearly dependent-on-temperature 
cosmological constant. The ground state's effect on long-distance 
gauge-mode propagation is the generation of 
mass in the off-Cartan directions (adjoint Higgs mechanism in unitary gauge).

Both phenomena, the generation of a finite energy density of the ground state and the 
emergence of mass, are described by the BPS saturated, 
classical dynamics of a spatially homogeneous, and adjoint 
scalar field $\phi$ and a pure-gauge 
configuration\footnote{Quantum and statistical fluctuations of the field $\phi$ are 
proven to be absent at the resolution $|\phi|$ \cite{Hofmann2005}.} $a_\mu^{bg}$. The reader may wonder 
why the macroscopic field $\phi$ is in the adjoint 
representation of the gauge group (after coarse-graining the gauge rotations of $\phi$ are 
only dependent on euclidean time $\tau$ \cite{Hofmann2005,HerbstHofmann2004}.) 
The answer to this question is rooted in the perturbative renormalizability of 
the Yang-Mills theory to any loop order \cite{'t HooftVeltmann} which 
guarantees that, after coarse-graining, propagating gauge fields of trivial topology 
are in the same representation 
as the gauge fields defining the fundamental action: Fundamental and 
coarse-grained gauge modes simply differ by a wave-function renormalization. 
A gauge-invariant coupling to the inert, coarse-grained sector 
with $|Q|=1$ thus is only possible by a covariant 
derivative for either a fundamental or an adjoint scalar. Since $\phi$ is a (nonlocal!) composite 
of the fundamental field strength the former possibility 
is ruled out (no spin-1/2 part in products of spin-1 representations). 

Technically, the coarse-graining over 
noninteracting\footnote{To assume a starting situation for the coarse-graining process, 
where (anti)calorons are noninteracting, turns out to be 
selfconsistent: The emerging, macroscopic field $\phi$ is inert, that is, not 
deformable by the microscopic interactions between (anti)calorons \cite{Hofmann2005}.} HS 
(anti)calorons is performed in two steps: (i) Determine the differential operator 
${\cal D}$ whose kernel ${\cal K}$ contains $\phi$'s phase $\hat{\phi}$ 
in terms of a unique definition of ${\cal K}$ involving a $\rho$ and an infinite space 
average over HS (anti)calorons\footnote{The dimensionless 
quantity $\hat{\phi}$ depends on temperature only via the periodicity in $\tau$. This implies  
that dimensional transmutation does not play a role in the emergence of $\hat{\phi}$. Thus 
$\hat{\phi}$ is obtained from an average over absolutely stable, 
classical field configurations: HS calorons. The definition of $\hat{\phi}$   
see \cite{Hofmann2005,HerbstHofmann2004}, excludes the contribution of BPS saturated 
configurations with $|Q|>1$. This is consistent 
with the final result that the field $\phi$ emerges due 
to a sufficiently local coarse graining down to a resolution 
$|\phi|$ such that multiple bumps of topological charge density 
do not enter the process. The integration over shifts of the (anti)caloron center is 
dimensionally forbidden. A fixed shift of the center w.r.t. the spatial origin 
can be related to the unshifted situation by an 
appropriate parallel transport which does not alter ${\cal D}$. 
The computation requires the introduction of various regularizations. A regularization 
of the azimuthal angular integration seems, at first sight, to break rotational 
invariance by the introduction of an axis. However, one can show 
that a rotation of this axis within the azimuthal plane is 
just a global gauge transformation of $\phi$'s phase. 
Thus any rotated axis would have yielded the same 
physics. As a consequence, no breaking of rotational symmetry is 
introduced by the regularization 
of the azimuthal angular integration.} entering in the definition of 
an adjoint two-point function of the field strength \cite{Hofmann2005,HerbstHofmann2004}, 
and (ii) assume the existence of a Yang-Mills scale 
$\Lambda$ to selfconsistently determine $|\phi|(T,\Lambda)$ by identifying the kernel 
${\cal K}^\prime$ of a `square-root' of ${\cal D}$ 
(BPS saturation, still noninteracting HS caloron and anticaloron). 
The resolution $|\phi|$ is optimal in the sense that 
it sets a length scale $|\phi|^{-1}$ at which the infinite 
spatial coarse-graining, used to define the phase $\hat{\phi}$, is saturated, and 
at the same time assures that the coarse-graining is sufficiently 
local such that configurations with $|Q|>1$ 
need not be taken into account. Namely, expressing the spatial ultraviolet 
cutoff $|\phi|^{-1}=\sqrt{\frac{2\pi}{\Lambda^3\beta}}$ in units of $\beta\equiv 1/T$, yields 
$8.22$ at the critical temperature $T_c$; for $T>T_c$ this number grows as $(T/T_c)^{3/2}$. 
But for integration cutoffs $\rho_u\sim r_u\ge 8.22\,\beta$ the kernel 
${\cal K}$ practically coincides with that of the infinite-volume limit, 
see \cite{Hofmann2005,HerbstHofmann2004}. Here $r_u$ 
denotes the infrared cutoff in the radial part of the space integral defining $\hat{\phi}$. 
Finally, one shows the above-mentioned fluctuation inertness of the field $\phi$ by 
means of its effective action whose potential is uniquely fixed\footnote{No freedom 
of shifting the ground-state energy density exists 
because of the BPS saturation of the field $\phi$.}.

Two parameters, the {\sl effective} coupling $e$ and the 
Yang-Mills scale $\Lambda$ (through the potential $V(\phi)$), enter 
the effective action. While the latter is a free parameter of the 
theory\footnote{Possibly derivable from the Planck mass $M_P$ 
if the emergence of gravity and matter will ever be understood 
in terms of a dynamical breaking of an infinite gauge symmetry.} 
the former is subject to an evolution 
in temperature. To derive the associated evolution 
equation we need to discuss the effects on the (quasiparticle) spectrum 
of the above a priori estimate for the ground state. 

First, 
one observes that a singular but admissible 
gauge transformation de-winds the field $\phi$ to lie in fixed direction on 
the group manifold (no $\tau$ dependence). At the same time 
$a^{bg}_\mu$ is gauged to zero (unitary gauge). As an aside one proves the deconfining nature of the 
discussed phase of the theory: The ground-state expectation of the 
Polyakov loop is, indeed, $Z_2$ degenerate. 
In unitary gauge the physical spectrum of excitations is identified: for SU(2) two out of three color 
directions acquire masses by the adjoint Higgs mechanism. We will refer 
to the massless direction as
tree-level massless (TLM) and to the massive one as tree-level heavy (TLH) (thermal quasiparticles). A completely physical 
gauge is reached by prescribing the Coulomb condition for the unbroken 
U(1)$\subset$SU(2). Thermodynamical quantities now 
are loop-expanded in terms of residual $Q=0$ 
fluctuations. That is, the process of 
integrating out these fluctuations is 
organized as a formal expansion in powers 
of $\hbar^{-1}$ and not in $e$. In unitary-Coulomb gauge there are 
conditions for the maximal off-shellness of fluctuations and for the maximal momentum 
transfer in a four-vertex, see Sec.\,\ref{CC}. Second, the invariance of Legendre transformations 
between thermodynamical quantities under the 
applied spatial coarse-graining is assured by the stationarity of the pressure 
with respect to variations in the Higgs-mechanism induced quasiparticle masses. 
This condition generates an equation for the evolution of $e$ with temperature. As a consequence, 
a decoupling of ultraviolet from infrared physics occurs. That is, $e$'s evolution at lower 
temperatures is independent of the physics taking place at 
a specifically chosen high temperature \cite{Hofmann2005}. 

The paper is organized as follows: In the next section 
we discuss and set up the constraints on the loop momenta for coarse-grained trivial-topology 
fluctuations as they emerge in the effective theory in unitary-Coulomb gauge. 
In Sec.\,\ref{PS} we argue that the resummation of 1PI contributions to 
the polarization tensor generates broadened 
spectral functions for each propagating mode and 
thus evades the problem of pinch singularities (powers of delta functions). 
In Sec.\,\ref{diagrammar} we generate a lower and an upper estimate for the 
ratio of the number of independent radial loop-momenta components to 
the number of constraints as a function of the number of 
vertices in bubble diagrams with resummed one-particle 
irreducible insertions. These estimates 
strongly suggest that only a finite number of such diagrams contribute to 
the loop expansion. An investigation of two 
example-diagrams is performed in Sec.\,\ref{examp}, 
one with a noncompact support for the radial loop-momenta 
integration and one where this integration is supported by a 
compact region. In Sec.\,\ref{Conc} we briefly summarize our 
results and give our conclusions.   

\section{Constraints on loop momenta\label{CC}}  

Here we discuss how and which constraints emerge for the propagation and interaction of coarse-grained, 
topologically trivial fluctuations in the effective theory. It is essential 
to note that the implementation of momentum 
constraints on residual fluctuations, arising due to a limited resolution, 
can only be performed if a real-time treatment of these fluctuations is applied. 
Only then is it possible to discern quantum from thermal fluctuations.   

The concept of a resolving power $R$ 
attached to a probe is based on the uncertainty relation:
\eqb
\label{uncrel}
\Delta x \Delta p\sim 1\,\ \ \ \ \ \Leftrightarrow\,\ \ \ \ \ \  \Delta p\sim \frac{1}{x}\equiv R\,.
\eqe
Here $\Delta p$ refers to the average {\sl deviation} of the probe momentum from 
the situation where the probe would move along a classical 
trajectory subject to a certain momentum distribution. In the case of a free 
theory $\Delta p$ is the deviation from the mass shell:
\eqb
\label{shellcond}
\Delta p=\sqrt{|p^2-m^2|}\,.
\eqe
Two sorts of observations are important: (i) In the effective theory, which is obtained by averaging over 
fluctuations with resolving power $\Delta p$ larger than the scale $|\phi|$, 
we need to consider only those modes with 
$\Delta p\le|\phi|$ for otherwise we would double-count 
propagating fluctuations. (ii) Since the process of coarse-graining generates particle 
masses only for $\Delta p\le |\phi|$ we need 
to make sure that an intermediate {\sl massless} particle in the fundamental theory, 
which as a massive particle does not exist in the effective theory and thus dresses a 
vertex such that the latter appears to be {\sl local}, does not possess a resolving power 
larger than $|\phi|$ in the fundamental theory for otherwise the vertex would no longer 
appear to be local\footnote{At $T=0$ and in pure perturbation theory 
this goes under the name renormalization at the scale $\mu$ where $\mu$ is determined 
by the maximal resolving power associated 
with the process under investigation. In the deconfining phase of 
Yang-Mills thermodynamics the resolving power $|\phi|$ selfconsistently 
emerges as a function of $T$ and $\Lambda$, see below and \cite{Hofmann2005}.}. 
Notice that the requirement $|p^2|\le|\phi|^2$ for the massless mode in the 
fundamental theory indeed implies that in the effective 
theory the massive mode with the same momentum is far away from 
its mass-shell and thus does not exist.

To work observations (i) and (ii) into 
quantitative constraints a completely fixed, physical gauge must 
be used. The absorption of the would-be 
Goldstone mode into the longitudinal component of the gauge field, emerging due to the apparent 
dynamical gauge-symmetry breaking SU(2)$\rightarrow$U(1), is facilitated 
by an admissible rotation to unitary gauge $\phi\equiv\lambda_3|\phi|$, 
$a_\mu^{g.s.}=0$ \cite{Hofmann2005}. In this gauge the spectral manifestation of the 
symmetry breaking is a quasiparticle mass for two out of three directions in 
the algebra. The remaining U(1) gauge freedom for fluctuations $\delta a^3_\mu$ is 
fixed by the Coulomb condition $\pd_i\delta a^3_i=0$.

Quantitatively, observation (i) then is expressed as
\eqb
\label{cond1}
|p^2-m^2|\le |\phi|^2\,\ \ \ (\mbox{for a TLH mode})\,,\ \ \ \ \ 
|p^2|\le |\phi|^2\,\ \ \ (\mbox{for a TLM mode})\,
\eqe
where $|\phi|=\sqrt{\frac{\Lambda^3}{2\pi T}}$. 
For a three-vertex (ii) is already contained in (i) 
by momentum conservation in the vertex. 
For a four-vertex the implementation of observation (ii) is 
more involved because one needs to distinguish $s$, $t$, and $u$ 
channels in the scattering process. Suppose that the ingoing (outgoing) momenta 
are labeled by $p_1$ and $p_2$ ($p_3$ and $p_4=p_1+p_2-p_3$). 
Then the following three conditions emerge
\eab
\label{cond2}
|(p_1+p_2)^2|&\le&|\phi|^2\,,\ \ \ (s\ \mbox{channel})\ \ \ \ \ \ \ \ \
|(p_3-p_1)^2|\le|\phi|^2\,,\ \ \ (t\ \mbox{channel})\nonumber\\ 
|(p_2-p_3)^2|&\le&|\phi|^2\,,\ \ \ (u\ \mbox{channel})\,.
\eae
Notice that the three conditions in Eq.\,(\ref{cond2}) reduce to 
the first condition if one computes the one-loop 
tadpole contribution to the polarization tensor or the two-loop contribution 
to a thermodynamical quantity, say the pressure, arising from a four 
vertex \cite{HerbstHofmannRohrer2004,SchwarzHofmannGiacosa2006BB}. Namely, the $t$-channel condition is then trivially satisfied 
while the $u$-channel condition reduces to the $s$-channel condition by letting 
the loop momentum $k\to -k$ in $|(p-k)^2|\le |\phi|^2$, 
see \cite{Hofmann2005,HerbstHofmannRohrer2004,SchwarzHofmannGiacosa2006BB}.  

Notice that upon a eulidean rotation $p_0\to ip_0$ 
the first condition in(\ref{cond1}) goes over in
\eqb
\label{cond1E}
|p^2+m^2|\le |\phi|^2\,.
\eqe
For SU(2) the quasiparticle mass is given as $m=2e\,|\phi|$ with $e\ge 8.89$ \cite{Hofmann2005}. 
Thus condition (\ref{cond1E}) is never satisfied, and 
TLH modes propagate on-shell only\footnote{Pairs of TLH modes cannot be created or 
annihilated by quantum processes because these would need to invoke 
momentum transfers of at least twice their mass. This, however, is about thirty five times 
larger than the maximally allowed resolving power in the effective theory. 
Due to the ground state possessing a positive energy density it is 
possible that a preexisting TLH mode of positive energy correlates 
with the oppositely charged negative-energy state 
(a localized depression of the energy density of the ground state) 
for a time interval of length larger than $|\phi|^{-1}$, see calculations in 
\cite{HerbstHofmannRohrer2004,SchwarzHofmannGiacosa2006BB}. Notice that this has nothing to do with 
the {\sl creation} and subsequent {\sl annihilation} of a pair of 
oppositely charged TLH modes.}. 

\section{Pinch singularities\label{PS}}

Here we would like to point out that the 
occurrence of powers of delta functions of the same 
argument, as they appear when real-time expanding the pressure into loops, 
can be resolved by appropriate resummations. 

\begin{figure}
\begin{center}
\leavevmode
\leavevmode
\vspace{4.3cm}
\includegraphics{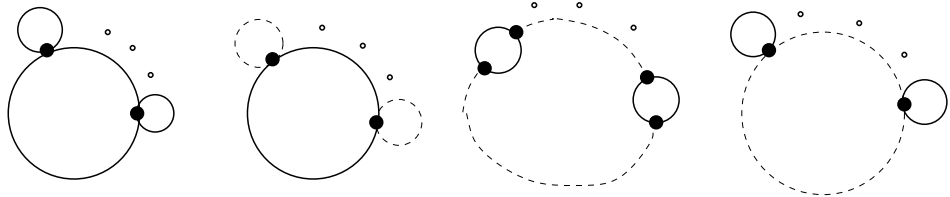}
\end{center}
\caption{\protect{\label{Fig-1}} Ring-diagrams as they occur in a loop expansion 
of the pressure in SU(2) Yang-Mills thermodynamics. Solid (dashed) 
lines are associated with TLH- (TLM-) mode propagation.}      
\end{figure}

The problem occurs in the 
so-called ring-diagrams, see Fig.\,\ref{Fig-1}.
At tree-level the propagators of a TLH or a TLM mode are given as 
\cite{SchwarzHofmannGiacosa2006BB}:
\begin{align}
\label{TLHprop}
D^{\tiny\mbox{TLH,0}}_{\mu\nu,ab}(p)&= -\delta_{ab}\tilde{D}_{\mu\nu}
\left[\frac{i}{p^2-m^2}+2\pi\delta(p^2-m^2)n_B(|p_0|/T) \right]\\
\tilde{D}_{\mu\nu} &= \left( g_{\mu\nu}-\frac{p_\mu p_\nu}{m^2} \right)
\end{align}
where $n_B(x)=1/(e^x-1)$ denotes the Bose-Einstein distribution function. 
For the free TLM mode we have
\begin{equation}
\label{TLMprop}
D^{\tiny\mbox{TLM,0}}_{ab,\mu\nu}(p) = -\delta_{ab}
\left\lbrace P^T_{\mu\nu}\left[\frac{i}{p^2}+2\pi\delta(p^2)n_B(|p_0|/T)\right]
-i\frac{u_\mu u_\nu}{\textbf{p}^2}\right\rbrace \,.
\end{equation} 
where 
\begin{align}
P^{00}_T & = P^{0i}_T = P^{i0}_T = 0\\
P^{ij}_T & = \delta^{ij} - p^{i}p^{j}/\textbf{p}^2\,.
\end{align}
TLM modes carry a color index 3 while TLH 
modes have a color index 1 and 2. Notice the term $\propto u_\mu u_\nu$ in 
Eq.\,(\ref{TLMprop}) describing the 'propagation' of the 
$A^3_0$ field. Here $u_\mu=(1,0,0,0)$ represents the four-velocity of the heat bath. 

Resumming one-particle irreducible (1PI) contributions to the polarization tensor, 
the scalar part of the tree-level propagators are modified 
in terms of screening functions $G_{TLH}(p)$ and $G_{TLM}(p)$ as\footnote{A discussion of the emergence of additional 
tensor structures $u_\mu u_\nu$ and $p_\mu u_\nu+p_\nu u_\mu$ due to loop effects is not important for our argument. 
To avoid a logical contradiction the one-loop polarizations are first computed in 
real time subject to the constraints (\ref{cond1}) and 
(\ref{cond2}). Subsequently, a continuation in the external momentum variable $p^0$ 
is performed to imaginary time. Then the resummation is carried out, and finally the result is 
continued back to real time.} 
\eab
\label{genprops}
&&\frac{i}{p^2-m^2}+2\pi\delta(p^2-m^2)\,n_B(|p_0|/T)\rightarrow\nonumber\\ 
&&\frac{i}{p^2-(m^2+\mbox{Re}\,G_{\tiny\mbox{TLH}}(p,T))}+2\pi\rho_{\tiny\mbox{TLH}}(p,T)\,n_B(|p_0|/T)\nonumber\\ 
&&\frac{i}{p^2}+2\pi\delta(p^2)\,n_B(|p_0|/T)\rightarrow\nonumber\\  
&&\frac{i}{p^2-\mbox{Re}\,G_{\tiny\mbox{TLM}}(p,T)}+2\pi\rho_{\tiny\mbox{TLM}}(p,T)\,n_B(|p_0|/T)\,,
\eae
where the spectral functions at fixed spatial momentum $\vec{p}$ are defined as
\eab
\label{imGdef}
\rho_{\tiny\mbox{TLH}}(p^0,\vec{p},T)&\equiv& \frac{1}{\pi}\,\mbox{Im}\,\frac{1}{p^2-(m^2+G_{\tiny\mbox{TLH}}(p,T))}\,\nonumber\\ 
&=&\frac{1}{\pi}\,\frac{\mbox{Im}\,G_{\tiny\mbox{TLH}}(p,T)}{\left(p^2-m^2-\mbox{Re}\,G_{\tiny\mbox{TLH}}(p,T)\right)^2+
\left(\mbox{Im}\,G_{\tiny\mbox{TLH}}(p,T)\right)^2}\,\nonumber\\ 
\rho_{\tiny\mbox{TLM}}(p^0,\vec{p},T)&\equiv& \frac{1}{\pi}\,\mbox{Im}\,\frac{1}{p^2-G_{\tiny\mbox{TLM}}(p,T)},\nonumber\\ 
&=&\frac{1}{\pi}\,\frac{\mbox{Im}\,G_{\tiny\mbox{TLM}}(p,T)}{\left(p^2-\mbox{Re}\,G_{\tiny\mbox{TLM}}(p,T)\right)^2+
\left(\mbox{Im}\,G_{\tiny\mbox{TLM}}(p,T)\right)^2}\,.
\eae
From Eq.\,(\ref{imGdef}) it 
follows that powers of $\delta$-functions of the same argument relax 
to powers of finite-widths (Lorentz-like) peaks 
of the same argument and thus to mathematically 
well-defined objects. The compositeness constraints and combinatorial factors are 
then modified compared with the tree-level ones. In practice, 
the computation of the pressure in a truncation at the two-loop level using tree-level 
propagators already yields results that are accurate on the 0.1\%-level 
\cite{HerbstHofmannRohrer2004,SchwarzHofmannGiacosa2006BB}.

\section{Connected bubble diagrams\label{diagrammar}} 

For the exact computation of thermodynamical quantities 
such as the pressure all diagrams contributing to 
each mode's full propagator need to be known. Knowing the exact 
propagator (or the polarization tensor), in turn,  fixes the exact dispersion law 
for each mode. This is important in applications \cite{SchwarzHofmannGiacosa2006BB}. 
The polarization tensor is a sum over connected bubble diagrams (loop diagrams with no external legs) 
with one internal line of momentum $p$ cut, such that the diagram reamins connected, 
and the two so-obtained external lines amputated. As a consequence, the vanishing of a connected 
bubble diagram due to a zero-measure support for its 
loop-momenta integrations implies that the associated 
contribution to a polarization tensor is also nil.   

We conjecture that {\sl all nonvanishing}, 
connected bubble diagrams enjoy the following property: 
For the total number $V$ of their vertices we have $V\le V_{\tiny{max}}$ with $V_{\tiny{max}}<\infty$ provided that all 
1PI contributions to the polarization tensor with up to $V_{\tiny{max}}$ many vertices are resummed. 

Let us now present our arguments in favor of this claim. The requirement 
that 1PI contributions to the polarization tensor are resummed assures that
(A) all vertex constraints in (\ref{cond2}) 
are operative (subject to slightly modified dispersion laws) 
and that (B) pinch singularities (powers of delta functions)
do not occur because of the broadening of the 
spectral function of the respective mode's propagator, 
compare with Sec.\,\ref{PS}. We consider the two 
cases where a connected bubble diagram {\sl solely} contains 
(i) $V_4$-many four-vertices and (ii) $V_3$-many three-vertices. 
This is relevant because the ratio of the number $\tilde{K}$ 
of independent radial loop-momentum variables (zero-components 
and moduli of spatial momenta) to the number $K$ of constraints on them is minimal 
for (i) and maximal for (ii) at a given number $V=V_3+V_4\ge 2$ of vertices, see Eq.\,(\ref{resultrat}). 
 
The relation between the number $L$ of independent loop momenta, 
the number $I$ of internal lines, and the number $V$ of vertices for planar bubble diagrams is \cite{WeinbergI}:
\eqb
\label{Eulercharc}
L=I-V+1\,.
\eqe
In the case (i) and (ii) we have in addition \cite{WeinbergI}
\eqb
\label{IV}
I=2\,V_4\,,\ \ \ \ \ \ \mbox{and}\ \ \ \ \ \ \ I=\frac{3}{2}\,V_3\,,
\eqe
respectively. According to (\ref{cond1}) we thus 
have in case (i) $ 2\,V_4$ constraints (propagators) and, according to (\ref{cond2}), 
at least $\frac{3}{2}\,V_4$ constraints (vertices) 
on loop momenta. In case (i) this gives a total of 
$K\ge \frac{7}{2}\,V_4$ constraints. In case (ii) we have a total of $K=\frac{3}{2}\,V_3$ 
constraints (only propagators). Combining Eqs.\,(\ref{Eulercharc}) and (\ref{IV}), 
we obtain: 
\eab
\label{zwres}
&&\mbox{for (i):}\ \ L=V_4+1\ \ \ \Rightarrow \ \ \ \tilde{K}=2\,V_4+2\,, \nonumber\\ 
&&\mbox{for (ii):}\ \ L=\frac{V_3}{2}+1\ \ \ \Rightarrow \ \ \ \tilde{K}=V_3+2\,.
\eae
This yields: 
\eqb
\label{resultrat}
\mbox{for (i):}\ \ \frac{\tilde{K}}{K}\le \frac{4}{7}\left(1+\frac{1}{V_4}\right)\,,\ \ \ \ \ 
\mbox{for (ii):}\ \ \frac{\tilde{K}}{K}=\frac{2}{3}\left(1+\frac{2}{V_3}\right)\,.
\eqe
Obviously, the ratio $\tilde{K}/K$ is smaller in case (i) than it is in case (ii). 
Notice that in case (i) the ratio $\frac{\tilde{K}}{K}$ is smaller than 
unity for $V_4\ge 2$ while this happens for $V_3\ge 6$ in case (ii). 

Now the constraints (\ref{cond1}) and (\ref{cond2}) are independent 
inequalities instead of independent equations and thus do not identify independent 
hypersurfaces\footnote{By independent 
hypersurfaces $H_i$ ($i=1,\cdots,h\le\tilde{K}$) in a $\tilde{K}$-dimensional
 euclidean space ${\bf R}^{\tilde{K}}$
we mean that in a whole environment $U$ 
of a point in their intersection $\bigcap_{i=1}^h H_i$ the 
normal vectors $\hat{n}_i$ to $H_i$ 
(computed somewhere on $U\cap H_i$) are linearly independent. 
If $h=\tilde{K}$ then it follows that $\bigcap_{i=1}^{\tilde{K}} H_i$ is a 
discrete set of points.} in a $\tilde{K}$-dimensional euclidean space 
${\bf R}^{\tilde{K}}$. 
Rather, the inequalities (\ref{cond1}) or (\ref{cond2}) `fatten' hypersurfaces that would be obtained by setting their 
right-hand sides equal to zero\footnote{The case of a 
TLH mode, where only the thermal, on-shell part of the 
propagator contributes with a $\delta$-function weight 
can be figured as the limit of a fat 
hypersurface subject to a regular weight 
acquiring zero width but now subject to a singular weight.}. As a consequence, the situation 
$\frac{\tilde{K}}{K}=1$ fixes a discrete set of 
compact regions $C_{\tilde{K}}$ in ${\bf R}^{\tilde{K}}$ rather than a discrete set 
of points. If $\frac{\tilde{K}}{K}$ is sufficiently smaller than unity, which should be the case 
for sufficiently large $V_4$ and/or $V_3$ 
according to Eq.\,(\ref{resultrat}), then the associated 
diagram does not contribute: Fat hypersurfaces, 
specified by the number $\kappa\equiv K-\tilde{K}$ of constraints not used 
up for the determination of $C_{\tilde{K}}$, should have an intersection $C_{\kappa}$ such that 
\eqb
\label{inadd}
C_{\kappa}\cap C_{\tilde{K}}=\emptyset\,\ \ \ \ \ (\kappa\gg 1)\,. 
\eqe
Notice that 
according to Eq.\,(\ref{resultrat}) 
\eqb
\label{kapparange}
\frac{3}{4}\,\tilde{K}\stackrel{>}\sim\kappa\stackrel{>}\sim\frac{1}{2}\,\tilde{K}\,,\ \ \ \ \ \ \ (\tilde{K}\gg 1)\,.
\eqe
Although it is not rigorously guaranteed that $C_{\kappa}\cap C_{\tilde{K}}=\emptyset$ with $\kappa$ ranging as 
in (\ref{kapparange}) this is, however, rather plausible. 

For completeness let us investigate the generalization of Eq.\,(\ref{Eulercharc}) to nonplanar 
bubble diagrams which can be considered spherical polyhedra with one face removed and a nonvanishing number 
of handles (genus $g>0$).  Eq.\,(\ref{Eulercharc}) then generalizes as 
\eqb
\label{gen}
V-I+L+1=2\ \ \longrightarrow \ \ V-I+L+1=2-2g\,,
\eqe
where again $I$ is the number of internal lines, $L$ the number of loops, and $g$ represents the genus of 
the polyhedral surface (the number of handles). Notice that the right-hand side of the right-hand 
side equation is the full Euler-L' Huilliers characteristics. Reasoning as above but now based on the 
general situation of $g\ge 0$ expressed by the right-hand side equation in Eqs.\,(\ref{gen}), we arrive at
\eab
\label{ratiosgen}
\frac{\tilde{K}}{K}&\le&\frac47\left(1+\frac{1}{V_4}\left(1-2g\right)\right)\,,\ \ \ (V=V_4)\,,\nonumber\\ 
\frac{\tilde{K}}{K}&\le&\frac23\left(1+\frac{2}{V_3}\left(1-2g\right)\right)\,,\ \ \ (V=V_3)\,.
\eae
According to Eqs.\,(\ref{ratiosgen}) the demand $\frac{\tilde{K}}{K}\le 1$ for a compact support of the 
loop integrations is always satisfied for $g\ge 1$ since the number of vertices needs to be positive: 
$V_4\ge 0$ and $V_3\ge 0$. Recall that at $g=0$ this is true 
only for $V_4\ge 2$ and $V_3\ge 6$, respectively. We thus conclude that bubble diagrams of a topology deviating from 
planarity are much more severely constrained than their planar counterparts.

\section{Two examples\label{examp}}

Here we would like to demonstrate how severely conditions (\ref{cond1}) and 
(\ref{cond2}) constrain the loop momenta when increasing 
the number of vertices. Namely, we compare the situation 
of a two-loop diagram with that of a three-loop diagram. 

\begin{figure}
\begin{center}
\leavevmode
\leavevmode
\vspace{4.3cm}
\includegraphics{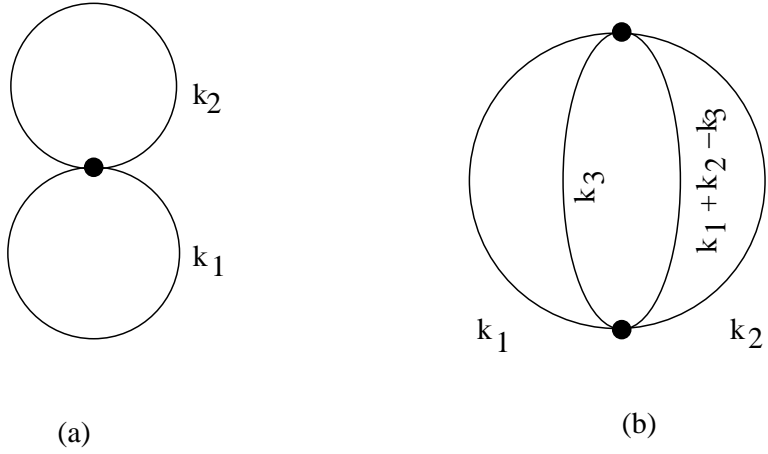}
\end{center}
\caption{\protect{\label{Fig-2}} (a) Two-loop and (b) three-loop diagram contributing 
to the pressure in the deconfining phase of SU(2) Yang-Mills thermodynamics. 
The solid lines are associated
with thermal TLH-mode propagation.}      
\end{figure}

Consider the diagrams in Fig.\,\ref{Fig-2}. Only TLH-modes are 
involved which, due to constraint (\ref{cond1E}), 
propagate thermally, that is, on their mass shell. Diagram (a) is real while diagram (b) 
is purely imaginary, but here we are only interested in their moduli.

\noindent{\sl Diagram (a):} We have $\tilde{K}=4$ and $K=3$. 
Thus the region of integration for the radial loop 
variables cannot be compact. Let us show this explicitly. Before applying the 
constraints in (\ref{cond2}) we have for diagram (a) 
\cite{HerbstHofmannRohrer2004}:
\eab
\label{Da}
\left|\Delta P_a\right|&=&\frac{e^2\,\Lambda^4\,\lambda^{-2}}{(2\pi)^4}\sum_{\pm}\int dx_1
\int dx_2 \int dz_{12}\,\frac{x_1^2\,x_2^2}{\sqrt{x_1^2+4e^2}\sqrt{x_2^2+4e^2}}\times\nonumber\\ 
&&P^{\pm}_a(x_1,x_2,z_{12})\,n_B\left(2\pi\lambda^{-3/2}\sqrt{x_1^2+4e^2}\right)\,
n_B\left(2\pi\lambda^{-3/2}\sqrt{x_2^2+4e^2}\right)\,,\nonumber\\ 
\eae
where $\lambda\equiv\frac{2\pi T}{\Lambda}$, 
$\vec{x}_i\equiv \frac{\vec{k}_i}{|\phi|}$, $x_{i}\equiv |\vec{x}_i|$ $(i=1,2)$, 
$z_{12}\equiv\cos\angle(\vec{x_1},\vec{x_2})$, 
and $P^{\pm}_a(x_1,x_2,z_{12})$ is given as:
\eab
\label{Pa2}
P^{\pm}_a(x_1,x_2,z_{12})&\equiv&\frac12\left(6-\frac{x_1^2}{4e^2}-\frac{x_2^2}{4e^2}-
\frac{x_1^2x_2^2}{16e^4}(1+z^2_{12})\pm \right.\nonumber\\ 
&&\left.2x_1x_2z_{12}\frac{\sqrt{x_1^2+4e^2}\sqrt{x_2^2+4e^2}}{16e^4}\right)\,.
\eae
Applying the constraint $|(k_1+k_2)^2|\le|\phi|^2$, see (\ref{cond2}), we have 
\eqb
\label{cond1rec}
\left|4e^2\pm\sqrt{x_1^2+4e^2}\sqrt{x_2^2+4e^2}-x_1x_2z_{12}\right|\le\frac{1}{2}\,.
\eqe
Only the minus sign is relevant ($e>\frac{1}{2\sqrt{2}}$) 
in (\ref{cond1rec}). Thus the expression within the absolute-value signs 
is strictly negative, and we have
\eqb
\label{z12}
z_{12}\le\frac{1}{x_1x_2}\left(4e^2-\sqrt{x_1^2+4e^2}\sqrt{x_2^2+4e^2}+\frac{1}{2}\right)\equiv g_{12}(x_1,x_2)\,.
\eqe
Notice that $\lim_{x_1,x_2\to\infty}g_{12}(x_1,x_2)=-1$. Apart from a small compact 
region, where $g_{12}(x_1,x_2)\ge 1$ and which includes the point $x_1=x_2=0$ in the $(x_1\ge 0,x_2\ge 0)$--quadrant,  
the admissible region of $x_1, x_2$-integration ($-1\le g_{12}(x_1,x_2)<1$) is an 
infinite strip bounded by the two functions
\eab
\label{2curves}
x_2^u(x_1)&=&\frac{x_1+8e^2+\sqrt{1+16e^2}\sqrt{x_1^2+4e^2}}{8e^2}\,,\nonumber\\ 
x_2^l(x_1)&=&\frac{x_1+8e^2-\sqrt{1+16e^2}\sqrt{x_1^2+4e^2}}{8e^2}\,.
\eae
We conclude that the integration region for radial loop momenta is not compact. 
Large $x_1$- and/or $x_2$- values are, however, Bose suppressed 
in Eq.\,(\ref{Da}), and the ratio 
$\frac{|\Delta P_a|}{P_{\mbox\tiny{1-loop}}}$, as a function of 
$\lambda$, is at most of order $10^{-5}$ \cite{HerbstHofmannRohrer2004}. 

\noindent{\sl Diagram (b):}  Here we have $\tilde{K}=6$ and $K=7$. According to the general arguments 
in Sec.\,\ref{diagrammar} we know that the admissible region of 
radial loop integration either is compact or empty. After a rescaling of 
the loop momenta conditions (\ref{cond2}) are recast into
\eab
\label{cond2b}
z_{12}&\le&\frac{1}{x_1x_2}\left(4e^2-\sqrt{x_1^2+4e^2}\sqrt{x_2^2+4e^2}+\frac{1}{2}\right)\equiv
g_{12}(x_1,x_2)\,,\nonumber\\ 
z_{13}&\ge&\frac{1}{x_1x_3}\left(-4e^2+\sqrt{x_1^2+4e^2}\sqrt{x_3^2+4e^2}-\frac{1}{2}\right)\equiv
g_{13}(x_1,x_3)\,,\nonumber\\ 
z_{23}&\ge&\frac{1}{x_2x_3}\left(-4e^2+\sqrt{x_2^2+4e^2}\sqrt{x_3^2+4e^2}-\frac{1}{2}\right)\equiv
g_{23}(x_2,x_3)\,,
\eae
where $z_{12}\equiv\cos\angle(\vec{x}_1,\vec{x}_2)$, $z_{13}\equiv\cos\angle(\vec{x}_1,\vec{x}_3)$, 
and $z_{23}\equiv\cos\angle(\vec{x}_2,\vec{x}_3)$. Notice that 
$\lim_{x_1,x_2\to\infty} g_{12}=-1=-\lim_{x_1,x_3\to\infty} g_{13}=
-\lim_{x_2,x_3\to\infty} g_{23}$. Before applying the constraints (\ref{cond2b}) 
diagram (b) reads 
\eab
\label{Db}
\left|\Delta P_b\right|&\le&\frac{e^4\Lambda^4\lambda^{-2}}{2^3(2\pi)^6}\sum^2_{l,m,n=1}\int dx_1\int dx_2\int
dx_3\int dz_{12}\int dz_{13}\int_{z_{23,l}}^{z_{23,u}} dz_{23}
\nonumber\\ 
&&\frac{1}{\sqrt{(1-z^2_{12})(1-z^2_{13})-(z_{23}-z_{12}z_{13})^2}}\frac{x_1^2x_2^2x_3^2}
{\sqrt{x_1^2+4e^2}\sqrt{x_2^2+4e^2}\sqrt{x_3^2+4e^2}}
\times\nonumber\\ 
&&\delta\left(4e^2+(-1)^{l+m}\sqrt{x_1^2+4e^2}\sqrt{x_2^2+4e^2}-
x_1x_2z_{12}-\right.\nonumber\\ 
&&\left.((-1)^{l+n}\sqrt{x_1^2+4e^2}\sqrt{x_3^2+4e^2}-
x_1x_3z_{13})-\right.\nonumber\\ 
&&\left.((-1)^{m+n}\sqrt{x_2^2+4e^2}\sqrt{x_3^2+4e^2}-
x_2x_3z_{23})\right)\times\nonumber\\ 
&&\left|P_b(\vec{x},\vec{z},l,m,n)\right|\,n_B\left(2\pi\lambda^{-3/2}\sqrt{x_1^2+4e^2}\right)\,
n_B\left(2\pi\lambda^{-3/2}\sqrt{x_2^2+4e^2}\right)\times\nonumber\\ 
&&n_B\left(2\pi\lambda^{-3/2}\sqrt{x_3^2+4e^2}\right)\times\nonumber\\ 
&&n_B\left(2\pi\lambda^{-3/2}\left|(-1)^l\sqrt{x_1^2+4e^2}+(-1)^m\sqrt{x_2^2+4e^2}
+(-1)^n\sqrt{x_3^2+4e^2}\right|\right)\,.\nonumber\\ 
\eae
where $P_b$ is a function of $\vec{x}\equiv(x_1, x_2, x_3)$ and $\vec{z}\equiv(z_{12}, z_{13},z_{23})$ 
emerging from Lorentz contractions and thus is regular at $\vec{x}=0$ (mass gap for TLH modes). 
In addition, we define:
\eqb
\label{zlu}
z_{23,u}\equiv\cos\left|\arccos z_{12}-\arccos z_{13}\right|\,,\ \ \ 
z_{23,l}\equiv\cos\left|\arccos z_{12}+\arccos z_{13}\right|\,.
\eqe
Let us now construct a useful compact 
embedding for the compact integration region in 
$x_1, x_2,$ and $x_3$. First, consider the case that 
$x_1, x_2, x_3\ge R>0$. If $R>1$ then conditions (\ref{cond2b}) and (\ref{zlu}) do conflict
\footnote{The truth of this and the following statements is easily checked numerically.}. The case that two out of the three variables 
$x_1, x_2,$ and $x_3$ are larger than 2 while the 
third one is smaller than 1 is excluded by virtue 
of (\ref{cond2b}) since two angular 
integrations would then have no support. In a similar way, the case that 
one variable is larger than 2 while the other two are smaller than 1 
is excluded by a vanishing support for two of the 
angular integrations. The situation that one variable is smaller than 1, another variable 
is in between 1 and 2, and the third variable is larger than 2 is excluded because 
one angular integration would then have no support. On the other hand, the situation that two variables 
are smaller than 1 while the third one is in between 1 and 2 is not excluded. This goes also for the case that 
two variables are in between 1 and 2 while the third one is smaller than 1. We conclude that the region of 
integration allowed by the conditions (\ref{cond2b}) and by 
Eq.\,(\ref{zlu}) is compact and bounded by a ball of radius 3 
which is centered at $\vec{x}=0$. Hence there is a qualitative 
difference with diagram (a) where the region of $x_1$-$x_2$ 
integration is noncompact. 

Solving $g(\vec{x},\vec{z})=0$ for $x_1$, where $g$ is the 
function defined by the argument of the delta function in Eq.\,(\ref{Db}), we have
\eqb
\label{x1} 
x_1=\frac{ac}{b^2-c^2}+(-1)^n\sqrt{\left(\frac{ac}{b^2-c^2}\right)^2-\frac{4e^2b^2-a^2}{b^2-c^2}}\,,
\eqe
where 
\eab
\label{defsabc}
c&\equiv& x_2z_{12}-x_3z_{13}\,,\ \ \ \  a\equiv(-1)^{m+n}\sqrt{x_2^2+4e^2}\sqrt{x_3^2+4e^2}-x_2x_3z_{23}-4e^2\,,\nonumber\\ 
b&\equiv&(-1)^l\sqrt{x_2^2+4e^2}+(-1)^{m+1}\sqrt{x_3^2+4e^2}\,.
\eae
We present numerical results for the temperature 
dependence of the estimate for the modulus of diagram (b) elsewhere \cite{KavianiHofmann2006}.

\section{Conclusions\label{Conc}}

We have discussed how constraints on loop momenta, which emerge 
in the effective theory for the deconfining phase 
of SU(2) Yang-Mills thermodynamics \cite{Hofmann2005},  
enforce a loop expansion with properties 
dissenting from those known in perturbation theory. 
Namely, we have argued that modulo 
1PI resummations there is only a finite number of connected 
bubble diagrams contributing to the expansion of thermodynamical 
quantities or, by cutting one internal line, to the expansion of 
the polarization tensor. Our arguments of Sec.\,\ref{diagrammar} 
apply equally well to the SU(3) case. Because the quasiparticle 
mass spectrum is slightly more involved there are mild 
modifications of Secs.\,\ref{CC}, \ref{PS}, and \ref{examp} when going from SU(2) 
to SU(3), see \cite{Hofmann2005}. 

The reason for the improved convergence properties of loop expansions in 
the effective theory is clear: The spatial coarse-graining over both topological 
and plane-wave fluctuations selfconsistenty generates both 
a natural resolving power as a function of $T$ and $\Lambda$ and quasiparticle masses on tree level. As a consequence, 
residual quantum fluctuations have a small action as compared to 
$\hbar$. 
        
\section*{Acknowledgments}
The author would like to acknowledge useful and stimulating conversations 
with Francesco Giacosa, Dariush Kaviani, Jan Pawlowski, and Markus Schwarz.

\baselineskip25pt
\end{document}